\documentclass[runningheads]{llncs}
\usepackage[T1]{fontenc}
\usepackage{graphicx,verbatim}

\usepackage{tikz}
\usepackage{blindtext}
\usepackage{layouts}
\usepackage{booktabs}
\usepackage{caption}
\usepackage{wrapfig}
\usepackage{algorithm2e}
\usepackage{tcolorbox}
\usepackage{amsmath}
\usepackage{graphicx}
\usepackage{hyperref}
\usepackage{subcaption}
\usepackage{bm}
\usepackage{svg}
\usepackage{subcaption}
\usepackage{amssymb}

\renewcommand{\epsilon}{\varepsilon}
\newcommand{\norm}[1]{\left\|#1\right\|}
\newcommand{\fastmri}{fastMRI\xspace}
\newcommand{\map}{mAP$_{0.25}$\xspace}
\newcommand{\SDR}{SDR\xspace}

\newcommand{\addcomment}[3]{{\color{#1}\textsuperscript{\textbf{#3:}}#2}}

\newif\ifsubmission
\submissiontrue 

\ifsubmission
    \newcommand{\niko}[1]{{\textcolor{black}{#1}}}
\else
    \newcommand{\niko}[1]{{\addcomment{teal}{#1}{}}}
\fi

\begin{document}

\title{Mind the Detail: Uncovering Clinically Relevant Image Details in Accelerated MRI with Semantically Diverse Reconstructions}
\titlerunning{Uncovering Clinically Relevant Details in Accelerated MRI}

\author{Jan Nikolas Morshuis\inst{1} \and Christian Schlarmann \inst{1} \and Thomas Küstner \inst{1,2} \and \\ Christian F. Baumgartner\inst{1,3}\textsuperscript{*} \and Matthias Hein\inst{1}\textsuperscript{*}}
\authorrunning{Jan Nikolas Morshuis et al.}
\institute{University of Tübingen, Germany \and University Hospital of Tübingen, Germany \and University of Lucerne, Switzerland \\
    \email{nikolas.morshuis@uni-tuebingen.de}\\
    \textsuperscript{*}Shared last authorship
}
    
\maketitle
\begin{abstract}
In recent years, accelerated MRI reconstruction based on deep learning has led to significant improvements in image quality with impressive results for high acceleration factors. However, from a clinical perspective image quality is only secondary; much more important is that all clinically relevant information is preserved in the reconstruction from heavily undersampled data. In this paper, we show that existing techniques, even when considering resampling for diffusion-based reconstruction, can fail to reconstruct small and rare pathologies, thus leading to potentially wrong diagnosis decisions (false negatives). To uncover the potentially missing clinical information we propose ``Semantically Diverse Reconstructions'' (\SDR), a method which, given an original reconstruction, generates novel reconstructions with enhanced semantic variability while all of them are fully consistent with the measured data. 
To evaluate \SDR automatically we train an object detector on the fastMRI+ dataset. We show that \SDR significantly reduces the chance of false-negative diagnoses (higher recall) and improves mean average precision compared to the original reconstructions. The code is available on \href{https://github.com/NikolasMorshuis/SDR}{https://github.com/NikolasMorshuis/SDR}.


\end{abstract}
\section{Introduction}

Accelerated MRI reconstruction from undersampled k-space data is an inverse problem with infinitely many solutions. However, it is common to reconstruct only a single solution, thereby ignoring potentially different yet still plausible reconstructions. Deep-learning-based approaches, including novel techniques based on diffusion models, have led to impressive reconstruction results in terms of image quality even for high acceleration factors. On the other hand, in clinical practice, some pathological tissue changes can be small and may occur rarely, making them potentially underrepresented in the training data. As a result, overconfident reconstruction models may favor more likely solutions of healthy tissue if consistent with the measurement data, potentially leading to missed pathologies. Missing clinically relevant pathologies in the reconstruction can be a major obstacle to the application of accelerated MRI in clinical practice.

\begin{figure}[t]
    \centering
    \includegraphics[width=0.99\linewidth]{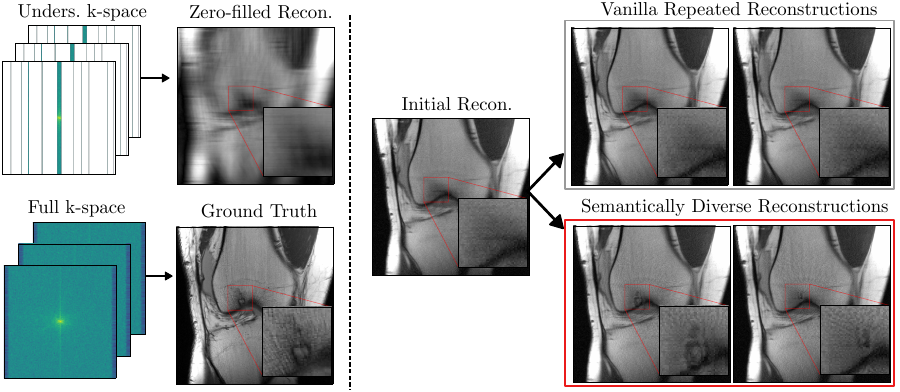}
    \caption{\textbf{Method overview.} At high acceleration factors (12x shown here), naive MRI reconstruction suffers from artifacts (left). Recent reconstruction methods like DDIP \cite{chung_ddip3d} can recover image details and generate multiple reconstructions, but these often lack diversity, potentially leading to missed pathologies (top right). Our proposed SDR approach produces semantically diverse reconstructions, helping detect pathologies that might otherwise be missed (bottom right).}
    \label{fig:teaser_fig}
\end{figure}

While some 
recent reconstruction methods such as diffusion-models \cite{chung2022score,chung_ddip3d,jalal2021robust} are capable of generating multiple different solutions by repeated sampling, it has been shown that these models tend to sample the most likely solutions frequently but only rarely sample less likely solutions that might still be diagnostically relevant~\cite{cohen2024from} (see Fig.\,\ref{fig:teaser_fig} top right). Moreover, sampling multiple reconstructions can be time-consuming as it requires repeating the denoising process. 

To address this problem, we propose a technique to generate semantically diverse reconstructions (\SDR) that are fully consistent with the measured k-space data. Our \SDR approach can be applied post-hoc to initial reconstructions obtained using any reconstruction method, producing a set of alternative solutions that preserve data consistency while potentially revealing pathologies not present in the original reconstruction. This offers the unique possibility to study a diverse set of data-consistent solutions and their inherent uncertainty. Our method generates these alternative solutions in $\approx 3$ seconds per image, which is faster than the sampling time of diffusion models that can require several minutes per image \cite{jalal2021robust}.

Our technique relies on identifying candidate regions in reconstructed images and 
modifying them to alter their semantic content while ensuring consistency with the measured k-space data. To generate semantically diverse reconstructions, we develop an adversarially robust version of the detection algorithm ViTDet \cite{li2022exploring}, leveraging techniques from~\cite{schlarmann2024robustclip} for the ViT-Backbone. We then use the features of candidate regions to maximize the semantic distance from the existing reconstructions (see Fig.\,\ref{fig:teaser_fig} bottom right).  This is motivated by the observation that the Euclidean distance in the feature space of adversarially robust vision encoders is a good semantic metric and that maximizing distance in this feature space leads to semantically meaningful changes \cite{croce2024adversarially}. We explore two versions of our method: i) an automatic selection of regions to be optimized using the proposal network of the object detector, ii) manual annotation of a candidate image region by the human user, e.g. a doctor.

In experiments on the fastMRI+ dataset \cite{zhao2022fastmri+}, we demonstrate that our method is able to produce a semantically diverse set of possible MRI reconstructions that capture more possible pathologies compared to alternative sampling techniques based on diffusion models~\cite{chung_ddip3d} and the E2E-Varnet~\cite{e2evarnet}. We quantitatively verify this by showing that pathology recall -- obtained with a separate Faster-RCNN detector \cite{ren2016faster} -- outperforms the baseline techniques, while mean average precision is not negatively affected. 

\section{Related Work}
%

The generation of multiple solutions in deep-learning-based MRI reconstructions is a relatively recent area of research. Morshuis et al. \cite{Morshuis_Segmentation} proposed a method to produce diverse MRI-re\-con\-struc\-tions corresponding to minimum and maximum segmentations in a downstream task. This approach estimates the segmentation uncertainty due to the ill-posed nature of the reconstruction problem.
Küstner et al. \cite{Kuestner_Rec_Unc} trained models with different initial seeds to generate multiple reconstructions, which were then used to estimate reconstruction uncertainty. They observed that uncertainty was highest in image regions containing pathologies.
Recent studies have demonstrated the strong performance of diffusion models in MR reconstruction~\cite{jalal2021robust,chung2022score,chung_ddip3d,miccai2022_diffusion_rec}. Apart from producing high-quality reconstructions, these techniques can also be used to generate multiple solutions by repeated sampling from the posterior distribution of images that are consistent with the measurement data. 

Only few works have focused on pathology detection in MRI~\cite{meniscustear_detection,yen2024adaptive}. Other work has used the pathology annotations provided by some datasets \cite{zhao2022fastmri+,desai2021skm} to compare reconstruction metrics in these relevant areas \cite{Kuestner_Rec_Unc,desai2021skm} or to show that measurement noise can lead to changes in reconstructions in these areas \cite{Morshuis2022}. To our knowledge, ours is the first work to consider pathology detection in accelerated MRI reconstruction utilizing multiple reconstructions.

\section{Method}

\subsection{Background and Notation}

The goal of MRI reconstruction is to find a reconstructed image $\boldsymbol{\hat{x}}$ that is consistent with the measured k-space data $\boldsymbol{y}$ subject to the following equation:
\begin{equation}
    \boldsymbol{y} = \boldsymbol{A}\boldsymbol{\hat{x}} \quad \text{ with }  \quad \bm{y} \in \mathbb{C}^m, \quad \bm{A} \in \mathbb{C}^{m \times n}, \quad \bm{\hat{x}} \in \mathbb{C}^n,
    \label{eq:inverse}
\end{equation}
where $\bm{A}=M\mathcal{F}S$ with the coil sensitivity maps $S$, the Fourier transform $\mathcal{F}$ and the undersampling mask $M$. MRI acquisitions are often accelerated by only measuring a fraction of the k-space \niko{data} $\boldsymbol{y}$. 

The fully sampled image, which would have been obtained by measuring the entire k-space,  contains a (possibly empty) set of pathologies denoted as $\mathcal{S} = \{s_1, s_2, ...\}$. However, due to image degradation caused by acceleration, a different set of pathologies, $\mathcal{S'} = \{s'_1, s'_2, ...\}$, is visible in the reconstructed image $\bm{\hat{x}}$. Ideally, $\mathcal{S}$ and $\mathcal{S'}$ should be identical to ensure accurate diagnosis. However, there are infinitely many reconstructions that satisfy Eq.\,\ref{eq:inverse} with no guarantee that they correctly reconstruct pathologies.  As a result, $\mathcal{S'}$ may fail to include all pathologies present in the patient,  particularly for high acceleration factors. 

In this work, we aim to generate a set of reconstructions, $\mathcal{X} = \{\bm{\hat{x}}^{(1)},  \ldots, \bm{\hat{x}}^{(N_{rec})}\}$, such that when considered jointly, the reconstructed images capture a set of pathologies $\mathcal{S'}$ that includes as many of the ground truth pathologies $\mathcal{S}$ as possible (few false negatives) while not hallucinating pathologies (few false positives).

\subsection{Generating Diverse Solutions Maximizing Semantic Distance}

Our key contribution is a method which, given an initial reconstruction $\bm{\hat{x}}^{(1)}$, can efficiently generate a set of reconstructions $\mathcal{X}$ that are semantically diverse in the regions where pathologies are likely to occur. 
To achieve this, we iteratively generate additional reconstructions $\bm{\hat{x}}^{(i)}$, $i=2,\ldots,N_{rec}$, by maximizing semantic distance $d$ to all previous reconstructions $\bm{\hat{x}}^{(j)}$ in an iterative procedure
\begin{equation}
    \bm{\hat{x}}^{(i)} \leftarrow \mathcal{DC} \bigg(P_{B(\bm{\hat{x}}^{(1)},r)}\Big(\bm{\hat{x}}^{(i)} + \eta \nabla_{\bm{\hat{x}}^{(i)}} \sum\nolimits_{j \neq i}d(\bm{\hat{x}}^{(i)},\bm{\hat{x}}^{(j)})\Big) \bigg),
\end{equation}
where $\eta$ is the step size, $\nabla_{\bm{\hat{x}}^{(i)}}$ is the gradient operator, $P_{B(\bm{\hat{x}}^{(1)},r)}$ is the projection onto the $\ell_2$-ball of radius $r$ around $\bm{\hat{x}}^{(1)}$, and $\mathcal{DC}$ is the projection operator mapping the reconstruction $\bm{\hat{x}}^{(i)}$ on the data-consistent space. We ensure data-consistency by replacing the k-space data with the original measured data at the locations where a measurement was acquired. To prevent potential artifacts from occurring, we run M-Step Conjugate Gradient before, minimizing the distance of the predicted k-space to the measurement. 
Furthermore, the projection $P_{B(\bm{\hat{x}}^{(1)},r)}$ ensures that $||\bm{\hat{x}}^{(i)}-\bm{\hat{x}}^{(1)}||_2 \leq r$, where $r$ is the allowed perturbation radius set to $r=3$ in our experiments. The value has been chosen, as it allows for meaningful perturbations while keeping the number of new false positives low.

To measure semantic distance between images, we utilize the features of a ViTDet detection network $f_\phi$ \cite{li2022exploring}, which we have finetuned on the fastMRI+ data. Relevant proposal bounding boxes $\mathcal{B} = \{b_1, b_2,...\}$ can be selected using methods described in Sec. \ref{sec:boxprop}.
Given the box-features $\bm{a}^{(i)} = f_{\phi}(\bm{\hat{x}}^{(i)}, b)$ we define 
\begin{equation}
d(\bm{\hat{x}}^{(i)},\bm{\hat{x}}^{(j)})  = \sum_{b\in \mathcal{B}}\norm{ f_{\phi}(\bm{\hat{x}}^{(i)}, b) - f_{\phi}(\bm{\hat{x}}^{(j)}, b)}_2.
    \label{eq:rec-loss}
\end{equation}

The proposed algorithm is summarized in Alg.\,\ref{alg:net}.

\begin{algorithm}[t]
\caption{Generating semantically diverse reconstructions by optimizing the distance between box features $\bm{a}^{(i)}$ obtained by the robust ViTDet.}
\label{alg:net}
 \SetAlgoLined
\KwIn{$\bm{y}$, $\bm{\hat{x}}^{(1)}$, $f_{\phi}$, $RPN$, $N_{opt}$, $N_{rec}$, $r$} 
\KwOut{$\mathcal{X} = \{\bm{\hat{x}}^{(1)}, ..., \bm{\hat{x}}^{(N_{rec})}\}$}
$\mathcal{B}=\{b_1, b_2,...\} \leftarrow RPN (\bm{\hat{x}}^{(1)})$ or manually \tcp*{Generate Proposal Boxes}
\For{$i \gets 1$ \textbf{to} $N_{rec}$}{
    $\bm{\epsilon}^{(i)} \sim \mathcal{N}(0, \sigma^2) \text{ with } 
    \bm{\epsilon}^{(1)}=0 $ \tcp*{Keep init. recon.}

    $\bm{\hat{x}}^{(i)} \gets P_{B(\bm{\hat{x}}^{(1)},r)}\big(\bm{\hat{x}}^{(1)}+\bm{\epsilon}^{(i)}\big)$, where $B(\bm{\hat{x}}^{(1)},r)=\{ \bm{x}\,|\, \|\bm{x}-\bm{\hat{x}}^{(1)}\|_2\leq r\}$,

    $\bm{\hat{x}}^{(i)} \gets \mathcal{DC}(\bm{\hat{x}}^{(i)})$  \tcp*{Data consistency}
    
    $\bm{a}^{(i)} \leftarrow f_{\phi}(\bm{\hat{x}}^{(i)}, \mathcal{B})$ \tcp*{Get box-features}
}
\For{$k \gets 0$ \textbf{to} $N_{opt}$}{
\For{$i\gets 2$ \textbf{to} $N_{rec}$}{

$d^{(i)} = \sum_{j\neq i} || \bm{a}^{(i)} - \bm{a}^{(j)} ||_2$ \tcp*{Calculate feature distances}

$\bm{\hat{x}}^{(i)} \gets P_{B(\bm{\hat{x}}^{(1)},r)}\big(\bm{\hat{x}}^{(i)} + \eta \nabla_{\bm{\hat{x}}^{(i)}} d^{(i)}\big)$

$\bm{\hat{x}}^{(i)} \gets \mathcal{DC}(\bm{\hat{x}}^{(i)})$ \tcp*{Data consistency}

$\bm{a}^{(i)} \leftarrow f_{\phi}(\bm{\hat{x}}^{(i)}, \mathcal{B})$ \tcp*{Get box-features}

}
}
\end{algorithm}

\subsection{Adversarially Robust Backbone}
\label{sec:robust_backbone}
Our application relies on gaining semantically meaningful gradients from the box-feature encoder $f_\phi$ in Eq.~\eqref{eq:rec-loss}. While gradients of deep learning models are generally not interpretable~\cite{mahendran2014understanding}, it is well-known that adversarially robust models possess semantically more meaningful gradients \cite{tsipras2019robustness,santurkar2019imagesynthesis,augustin2020ratio,ganz2024clipag,croce2024adversarially}.
To obtain an adversarially robust detection model, we leverage the recently proposed adversarial fine-tuning method of \cite{schlarmann2024robustclip}. We apply this method to the ViT-backbone $g_{\phi}$ of the ViTDet. The ViT-backbone also serves as the backbone of the box-feature encoder $f_\phi = h_\phi \circ g_\phi$. This ensures the stability of both the backbone and box features under small adversarial perturbations. The training objective is
\begin{equation}\label{eq:robust-backbone}
   \min_\phi \sum_i \max_{\norm{\bm{\delta}}_2 \leq r} \norm{g_\phi(\bm{x}_i+\bm{\delta}) - g_{\phi_\textrm{org}}(\bm{x}_i)}_2^2,
\end{equation}
where $g_{\phi_\textrm{org}}$ is the frozen original backbone,  $r$ is the adversarial perturbation budget, and $\mathbf{x}_i$ are unlabeled training images. In contrast to \cite{schlarmann2024robustclip}, we use the $\ell_2$-norm instead of the $\ell_\infty$-norm to bound the perturbation budget, as this generally yields better generative capabilities~\cite{augustin2020ratio}.
\begin{figure}[t]
    \centering
    \includegraphics[width=0.99\linewidth]{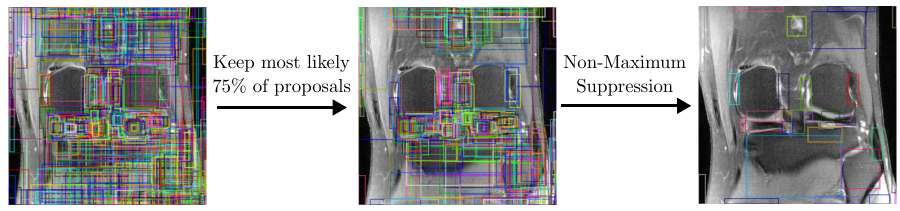}
    \caption{\textbf{Automatic Proposal Generation}. Given an initial reconstruction, we filter the proposal boxes given by the Region Proposal Network (RPN) of the ViTDet to only keep the 75\% of boxes with the highest score. We then apply Non-Maximum Suppression (NMS) on the remaining proposals. The remaining proposal boxes are then used for optimization with \SDR-A.}
    \label{fig:proposal_box_generation}
\end{figure}
We set $r$ to 10 and train for 20,000 steps on ImageNet data at resolution 640 using a batch size of 16. While it would be natural to train on \fastmri images, we have observed a stronger performance when training on a much larger set of diverse natural images.

\subsection{Box-Proposal Selection}
\label{sec:boxprop}

We introduce two approaches for identifying the box regions to optimize, which are motivated by two different use-cases. 

With \textbf{Manual (SDR-M) Proposal Box Generation} we address the scenario where a user suspects a pathology in a specific area and manually draws a bounding box. We simulate user annotations by perturbing the ground-truth boxes in size (75\%-125\%) and position (up to 25\% of width or height). 

With \textbf{Automatic (SDR-A) Proposal Box Generation} we address the case when no patient-specific information on the location of pathologies is available. In this scenario, we use the Region Proposal Network (RPN) of the ViTDet. We filter out the 25\% of boxes that have the lowest probability, as these are often in the background of the MRI image. We then utilize Non-Maximum-Suppression (NMS) to remove boxes with significant overlap (IoU > 0.05), see Figure \ref{fig:proposal_box_generation}.

After applying either method, we get a set of proposal boxes $\mathcal{B}=\{b_1, b_2, ...\}$ in which we optimize the reconstructions to be semantically different.

\begin{figure}[t]
    \centering
    \includegraphics[width=0.99\linewidth]{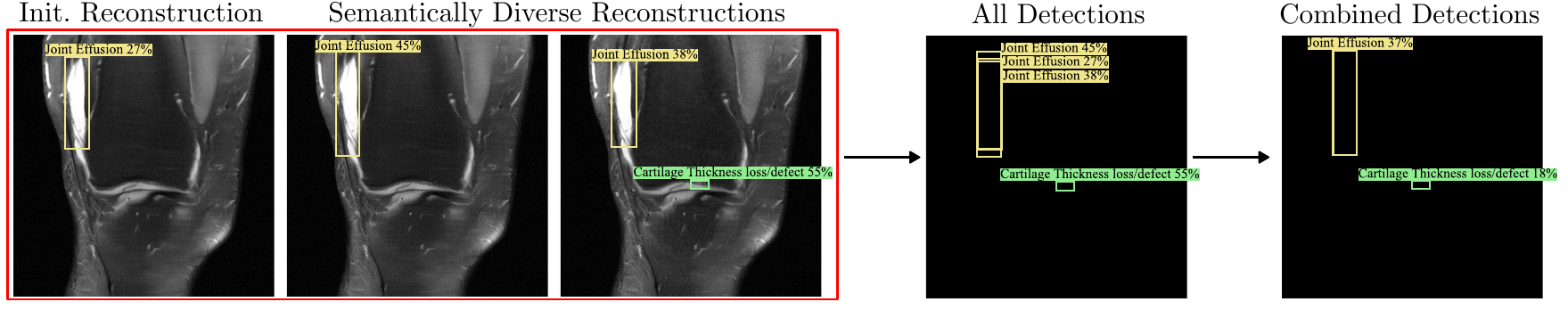}
    \caption{\textbf{Combining predictions of multiple reconstructions.} We combine overlapping detection boxes and average their probability score across reconstructed images to compute the mAP of our detections.}
    \label{fig:enter-label}
\end{figure}
\begin{figure}[t!]
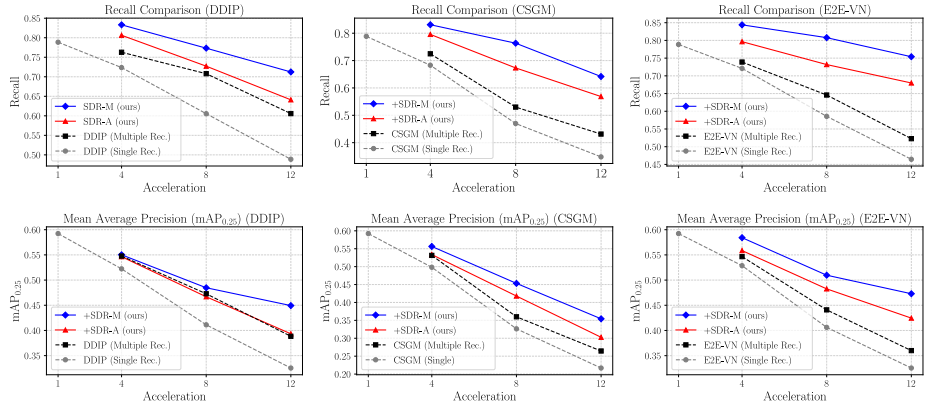

    \centering
    \begin{tabular}{ccc}
    \hspace{-1mm}\includesvg[width=0.33\textwidth]{lineplots/ddip_recall.svg} 
    & \includesvg[width=0.33\textwidth]{lineplots/csgm_recall.svg}
    &
    \includesvg[width=0.33\textwidth]{lineplots/e2e_recall.svg}\\
\hspace{-1mm}\includesvg[width=0.33\textwidth]{lineplots/ddip_map.svg} 
& \includesvg[width=0.33\textwidth]{lineplots/csgm_map.svg} &
    \includesvg[width=0.33\textwidth]{lineplots/e2e_map.svg}
    \end{tabular}
\caption{\textbf{Evaluation of reconstruction methods.} 
    Both SDR-A and SDR-M outperform single and multiple reconstructions using the original reconstruction techniques (DDIP, CSGM, E2E-VN) in terms of recall (better detection of pathologies) while achieving similar or better mAP$_{0.25}$.
    Thus SDR generates clinically relevant reconstructions outperforming resampling.
    }
    \label{fig:csgm_results_graph}
\end{figure}
\section{Data, Evaluation and Results}
Assessing clinical relevance of reconstructions would ideally require input from clinicians. As it is difficult to do such an evaluation at scale, instead, we evaluate by training an object detector and use it as a proxy for human annotators. \\
\textbf{Data and Training:} 
Open datasets of MRI reconstruction data are rare~\cite{fastmri,desai2021skm,solomon2024fastmri,tibrewala2024fastmri} and only a few provide annotated pathologies \cite{zhao2022fastmri+,desai2021skm}. For our experiments we use the fastMRI+ \cite{zhao2022fastmri+} knee data, as it provides enough annotated slices to train large detection networks. We split the fastMRI+ validation set subject-wise into test and validation set.
For evaluation, we considered the 10 most frequent pathology classes in fastMRI+, as the remaining classes have not enough
test samples.\\
\textbf{Object Detector for Evaluation:} To ensure an objective evaluation independent of the ViTDet used for generating the reconstructions of \SDR, we trained a separate Faster-RCNN detection network \cite{ren2016faster} which we use for evaluation.\\
\textbf{Combining detections of multiple images:}
Given a set of potential pathology detections $\mathcal{S}' = \{s'_1, s'_2, ...\}$ that was extracted from a set of reconstructions $\mathcal{X}$, we need to remove near duplicates and estimate the probability of each bounding box for the  calculation of mean average precision (mAP).
To remove near duplicates, we used NMS to eliminate overlapping bounding boxes (Intersection-over-Union (IoU) $\geq 0.25$) of the same class. To estimate the probability scores of the remaining boxes, we averaged the probability for a given bounding box $s'_i$ across all reconstructed images and set the probability score to 0 for reconstructions that did not contain the bounding box.\\
\textbf{Reconstruction Techniques:}
We analyzed the benefits of \SDR with three base reconstruction methods: the diffusion-based DDIP \cite{chung_ddip3d}, and CSGM \cite{jalal2021robust} approaches, as well as the E2E-VarNet~\cite{e2evarnet}. We produce 3 total reconstructions for each image using the baselines as well as \SDR, i.e. \SDR ($N_{rec}=3$). While the diffusion models can generate multiple reconstructions by design, we trained three separate E2E-VarNets with three different random seeds for each acceleration to reconstruct three different images during inference.\\
\textbf{Performance Measures:}
We measured the methods' ability to produce sets of reconstructions that contain the ground truth pathologies using the recall, where a recall of 1 indicates no missed pathologies. As more reconstructions might lead to more false positives, we also measured mean Average Precision (\map) \cite{everingham2010pascal}. For both metrics, we consider an IoU of 0.25 as correct detection. Small IoU thresholds are commonly used in medical detection tasks \cite{nndet} to account for uncertain border definitions and the typically small size of pathologies.

\begin{figure}[t!]
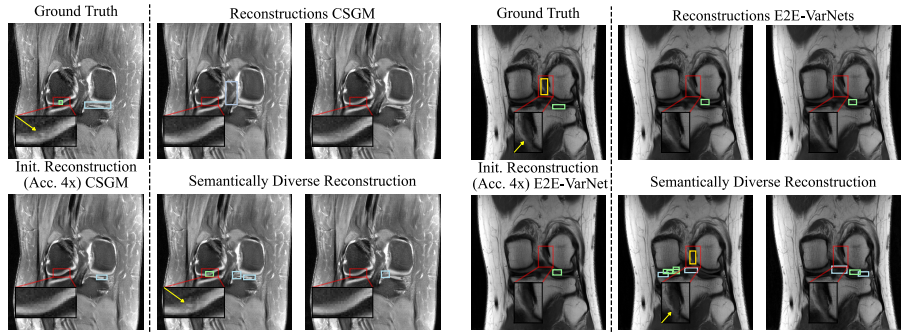

    \centering
    \begin{subfigure}{0.49\textwidth}
        \centering
        \includesvg[width=\textwidth]{plots/file1001440_025_cartiladge_csgm4x_arrows.svg}
    \end{subfigure}
    \hfill
    \begin{subfigure}{0.49\textwidth}
        \centering
        \includesvg[width=\textwidth]{plot_examples/example_rupture_arrow.svg}
    \end{subfigure}
    \hfill
    \caption{\textbf{Qualitative Results} 
    The highlighted box in the ground truth shows a cartilage thickness loss (left) and a high grade ligament sprain (right).  The initial reconstructions are often too smooth, preventing the detection of the pathologies. The semantic changes of \SDR-M (left) and \SDR-A (right) reveal the pathological changes. Yellow arrows indicate pathologies.
    }
\label{fig:qualitative_results}
\end{figure}

\section{Results}
The quantitative results in Fig.~\ref{fig:csgm_results_graph} show that \SDR-A and \SDR-M have higher recall than the original single and multiple reconstructions. Thus \SDR allows to uncover pathologies which can be missed with the original reconstruction methods. This is true for the diffusion-based reconstruction methods CSGM and DDIP as well as the often applied technique E2E-VarNet. At the same time \SDR even improves mAP$_{0.25}$ and thus \SDR is not simply achieving higher recall by introducing more false positive detections. 
We observed that it is necessary to use a ViTDet with robust backbone for the generation in SDR, as the original ViTDet did not improve in recall or mAP compared to repeated sampling.

\section{Discussion and Conclusion}
We have shown that existing accelerated MRI reconstruction methods can miss pathologies and have suggested \SDR to explore semantically different but fully data consistent reconstructions. The evaluation of \SDR reconstructions with an object detector shows fewer missed pathologies (higher recall)
while improving mAP. We believe that \SDR can lead to a safer and more reliable development of image reconstructions for accelerated MRI. 

The generated semantically diverse reconstructions can also be utilized by a clinician to better understand the reconstruction uncertainty, as the diverse reconstructions visualize disparities between the solutions. This would allow the clinician to acquire further measurements if necessary.

\begin{credits}
\subsubsection{\ackname} Funded by the Deutsche Forschungsgemeinschaft (DFG, Ger-
man Research Foundation) under Germany’s Excellence Strategy - EXC number 2064/1
- Project number 390727645. The authors thank the International Max Planck Research
School for Intelligent Systems (IMPRS-IS) for supporting JNM and CS.

\end{credits}

\bibliographystyle{splncs04}
\bibliography{main}

\end{document}